\documentclass[12pt]{article} \usepackage{cite,amsfonts,amssymb}
\newcommand{\nc}{\newcommand}
\nc{\beq}{\begin{equation}} \nc{\eeq}{\end{equation}}
\nc{\beqa}{\begin{eqnarray}} \nc{\eeqa}{\end{eqnarray}}
\nc{\lsim}{\begin{array}{c}\,\sim\vspace{-21pt}\\< \end{array}}
\nc{\gsim}{\begin{array}{c}\sim\vspace{-21pt}\\> \end{array}}
\def\Dslash{\not{\hbox{\kern-3pt $D$}}}

\begin{document}

\title{
{\baselineskip -.2in
\vbox{\small\hskip 4in \hbox{IITM/PH/TH/2007/2}}
\vbox{\small\hskip 4in \hbox{TIFR/TH/07-10~~~~~~~~}}} 
\vskip .4in
\vbox{
{\bf \LARGE On The Stability of Non-Supersymmetric Attractors in String Theory}
}
\author{Suresh Nampuri${}^1$\thanks{email:suresh@theory.tifr.res.in},~ 
Prasanta K. Tripathy${}^2$\thanks{email: prasanta@physics.iitm.ac.in}
~and
Sandip P. Trivedi${}^1$\thanks{email: sandip@theory.tifr.res.in} \\
\normalsize{\it ${}^1$ Department of Theoretical Physics,}\\
\normalsize{\it Tata Institute of Fundamental Research,} \\
\normalsize{\it Homi Bhabha Road, Mumbai 400 005, India.} \\
\normalsize{\it ${}^2$ Department of Physics,}\\
\normalsize{\it Indian Institute of Technology Madras,} \\
\normalsize{\it Chennai 600 036, India.}
}}
\maketitle
\begin{abstract}

We study non-supersymmetric attractors  obtained  in Type IIA compactifications on Calabi Yau manifolds. 
Determining  if an  
 attractor is stable or unstable requires an algebraically complicated analysis in general.
We show  using group theoretic techniques that   this analysis can be considerably 
simplified and  can be   reduced to solving a simple example like the STU model. 
For  attractors with  $D0-D4$ brane charges, determining stability requires expanding 
 the effective potential  to quartic order in the massless fields.  
We obtain the full set of these terms. 
For attractors with   $D0-D6$ brane charges,  we find that there is a moduli space of solutions
 and the resulting attractors are stable. 
Our analysis is restricted to the two derivative action.

\end{abstract}

\newpage
\section{Introduction and Overview}

Supersymmetric black holes are well known to exhibit the attractor phenomenon. This was first observed in 
\cite{Ferrara:1995ih} and subsequently explored in  several papers, for example, 
\cite{Strominger:1996kf,Ferrara:1996dd,Ferrara:1996um,Gibbons:1996af,Ferrara:1997tw, Denefa,Denef:2001xn}. 
More recently the study of non-supersymmetric extremal black holes has gained prominence and it has been realised
that these black holes are also attractors. Early investigations were carried out in, 
\cite{Gibbons:1996af,Ferrara:1997tw}.  Sufficient conditions for the existence of  a stable attractor
were discussed in terms of an effective potential in  \cite{Goldstein:2005hq}. A 
black hole entropy function was formulated which allows higher derivative corrections to also be 
included in \cite{senwa}.  
Subsequent investigations include \cite{Kraus:2005zm,kalloshb,Kallosh:2005ax,Tripathy:2005qp,Giryavets:2005nf,
Goldstein:2005rr,Alishahiha:2006ke,Chandrasekhar:2006kx,Bellucci:2006ew,Parvizi:2006uz,Kallosh:2006bx,
Sahoo:2006rp,Ferrara:2006em,Alishahiha:2006jd,Ferrara:2006yb,Astefanesei:2006dd,Kallosh:2006ib,Kaura:2006mv,
Cardoso:2006cb,Morales:2006gm,Bellucci:2006ib,Astefanesei:2006sy,Dabholkar:2006tb,Chandrasekhar:2006ic,
D'Auria:2007ev,Cai:2007ik,Ceresole:2007wx,Andrianopoli:2007rm,Saraikin:2007jc, Garousi:2007zb, Garousi:2007nn,
Cho:2007mn, Ferrara:2007}.

In particular, of particular relevance to this note, is the investigation carried out in \cite{Tripathy:2005qp},
where non-supersymmetric extremal black holes in Type IIA string theory compactified on a 
Calabi-Yau three-fold were analysed. An attractor corresponds to a critical point of the effective potential. 
For the attractor to be stable,  the  critical point must be a minimum of the effective potential.
Flat directions are allowed, but the extremum cannot be a maximum along any  direction of moduli space.
For the non-supersymmetric extrema found in \cite{Tripathy:2005qp}, it was noticed that the 
mass matrix, which governs the quadratic fluctuations about the extremum, has zero eigen values \footnote{In general hypermultiplets are not sourced by the gauge fields and will be flat directions of the effective potential in the two derivative approximation. The massless fields referred to here arise from the vector multiplets and can be  lifted by corrections beyond the quadratic order even in the two derivative theory.}. 
Thus it is necessary to expand the effective potential beyond quadratic order to determine
 whether the extrema are minima or maxima along the massless directions, or whether 
these directions are exactly flat.  
While  straightforward in principle, this procedure is algebraically quite involved in practice. 
The purpose of this note is to show that the required algebraic manipulations can be considerably simplified
by using   group theoretic considerations.  We find that these considerations determine  the general form of the 
terms which can arise
at the cubic, quartic etc levels  upto coefficients. The coefficients can then be determined by 
carrying out the calculations in a simple model like the STU model. 

This note is organised as follows. After some preliminaries, we first illustrate the above procedure by 
considering the case of a black hole which carries $D0,D4$ -brane charges and show how the mass matrix and 
quartic terms can  be determined  by a comparison with the STU case. This analysis also extends quite directly
to the case of a black hole carrying, $D0,D2,D4$ charges.  
Next,  we consider black holes which carry  $D6$ brane charge. We show in general that there is a moduli space of 
solutions in the D0-D6 system and the resulting attractor is stable. 
For the $D0-D4-D6$ case we argue that no terms cubic in the fluctuations can appear,
again by a comparison with the STU case. Throughout we work in the two derivative approximation,
so our results are applicable to ``big'' black holes. 

Our  analysis corrects some errors in earlier work in   \cite{Tripathy:2005qp}.
In particular the quadratic terms were not completely determined therein and it was incorrectly argued that  
 for the $D0-D4-D6$ case a cubic term is present which  renders the attractor unstable. 

There are several open questions worth investigating further. 
The quartic correction  in the $D0-D4$ case - which is the first correction along the massless directions - has an 
interesting structure and consists of  two terms with  opposite signs. Both terms depend on the triple intersection
numbers of the Calabi Yau manifold and the charges carried by the black hole. It is interesting to 
explore whether  the  stable or unstable nature of the  attractor can vary as this data is changed. 
We briefly discuss a model which illustrates this possibility towards the end of our discussion of the 
$D0-D4$ system.  
In the more general case, of a black hole which carries 
$D0-D4-D6$ charge, symmetries again allow the same two terms  at the quartic order.  
It should again  be a straightforward exercise to determine the coefficients by comparing, say against the STU model,
but we have not done so here. 
Finally, an interesting general issue is the inclusion of higher derivative corrections and how they 
alter the required conditions for the existence of a stable attractor. These terms should  be particularly important 
for the flat directions of the two-derivative effective potential and for the directions which are not
 flat but which have vanishing quadratic terms.

\section{Some Preliminaries}
Type IIA string theory compactified on a $CY3$ has ${\cal N}=2$ supersymmetry. Moduli in 
the resulting low-energy theory  lie in vector multiplets and hypermultiplets. The vector multiplet moduli 
couple to the gauge fields and are fixed by the attractor mechanism. The low-energy dynamics for the vector 
multiplets is determined by a prepotential.  If the Calabi-Yau manifold  has $h(1,1)=N$,
 there are $N$ vector multiplets and $N+1$ gauge fields in the low-energy theory. The prepotential 
(neglecting any $\alpha'$ corrections)
 is,
\begin{equation}
\label{prepot}
F=D_{abc}{X^aX^bX^c  \over X^0}.
\end{equation}
Here $X^0, X^a, a=1, \cdots N$ are the projective coordinates of special geometry, and 
$D_{abc}$ are the triple intersection numbers. 

The Kahler potential is given by, 
\beq
\label{defk}
K=-\ln\left[ i \sum_{A=0}^N \left(X^{A*}\partial_A F - X^A (\partial_A F)^{*}\right)\right].
\eeq
We will use the notation, 
\beq
\label{defx}
x^a={X^a \over X^0},
\eeq
in what follows,    
and use the projective invariance to set $X^0=1$. 
This gives a Kahler potential, 
\beq
\label{newk}
K = - \ln\left(-i D_{abc} (x^a - \bar x^a) (x^b - \bar x^b) (x^c - \bar x^c) \right) ~.
\eeq
A superpotential can be defined, it depends on the charges carried by the black hole. 
Let  $\Sigma^a, a=1, \cdots N$ and ${\hat \Sigma}_a$ be  
a basis of 4-cycles  and 2-cycles of the $CY_3$, and consider a 
 general black hole  carrying $(q_0,q_a, p^a, p^0)$ units of 
$D0-D2-D4-D6$ brane charge. The superpotential is then given by, 
\beq
\label{sp}
W=q_0 X^0 + q_aX^a  -p^a \partial_a F -p^0 \partial_0F ~.
\eeq
In the gauge $X^0=1$ this becomes, 
\beq
\label{spt}
W=q_0 + q_ax^a - 3 D_{abc}x^a x^b p^c + p^0 D_{abc}x^ax^bx^c ~.
\eeq
In the discussion which follows, we use the notation, 
\beq
\label{defD}
D_{ab}\equiv D_{abc}p^c, \ \ D_{a}\equiv D_{abc}p^bp^c, \ \ D\equiv D_{abc}p^ap^bp^c.
\eeq

The effective potential, which determines the existence of an attractor is given in terms of the superpotential and the Kahler potential by, 
\beq
\label{efpot}
V_{\rm eff}=e^K \left[ g^{a{\bar b}}\nabla_a W (\nabla_b W)^* + |W|^2 \right] ~, 
\eeq
where $g_{a {\bar b}} = \partial_a\partial_{\bar{b}}K$, $g^{a{\bar b}}$ is the inverse of $g_{a{\bar b}}$ and 
$\nabla_a W =\partial_aW + \partial_a K W$. 

For an attractor to exists $V_{\rm eff}$ must have an extremum. If this extremum is a minimum, the attractor is stable. 
The extrema of this effective potential were analysed in \cite{Tripathy:2005qp}. 
For the $D0-D2-D4$ system  the extremum is given at $x^a=x^a_0$ where  by, 
\beq 
\label{susyex}
x^a_0=ip^a \sqrt{{\hat q_0}\over D} + {1\over 6} D^{ab}q_b ~,
\eeq
in the supersymmetric case and,
\beq
\label{nonsusyex}
x^a_0=ip^a\sqrt{-{\hat q_0} \over D} + {1\over 6} D^{ab}q_b ~,
\eeq 
in the non-susy case. Here,
\beq
\label{defhatq0}
{\hat q_0}\equiv q_0+{1\over 12} D_{ab}q^aq^b~.
\eeq
The entropy of the non-supersymmetric extremal black hole is,
\beq
\label{enta}
S=2 \pi \sqrt{-D {\hat q_0}} ~.
\eeq
For the $D0-D4-D6$ system the non-susy extremum is given by,
\beq
\label{extgen}
x_0^a=p^a (t_1+ i t_2),
\eeq
where $t_1,t_2$ are determined by the charges and given in eq.(63,64) of section 3.3 in  \cite{Tripathy:2005qp}. 
The entropy of the non-supersymmetric extremal  black hole is,
\beq
\label{entb}
S=\pi\sqrt{(p^0)^2q_0^2-4Dq_0}.
\eeq

To determine whether the attractor is stable we now expand about the extremum. Let us define 
the fluctuations, $\delta \xi^a, \delta y^a$, as 
\beq
\label{defdelta}
x^a=x_0^a+ \delta x^a \equiv x_0^a + \delta   \xi^a + i \delta y^a.  
\eeq

For the 
 $D0-D4-D6$ system, with no $D2$-brane charge, 
the terms quadratic in the fluctuations take the form \footnote{When the $D6$-brane charge vanishes the $D2$-brane charge can be included in a straightforward manner}, 
\begin{eqnarray}
S_{\rm quadr}&=&\partial_a\partial_{\bar d}V(\delta\xi^{a}\delta\xi^{d}+\delta y^{a}\delta y^{d}) \cr
&& +Re(\partial_a\partial_dV)
(\delta\xi^{a}\delta\xi^{d}-\delta y^{a} \delta y^{d}) -2 Im(\partial_a\partial_dV)\delta\xi^{a}\delta\xi^{d}.
\end{eqnarray} 
(Note this corrects some typos of factors of two in eq.(117) of \cite{Tripathy:2005qp}). 

The mass matrix can then be read off and takes the form \footnote{Our conventions are that the 
quadratic terms are given by,
$$S_{\rm quadr}={1\over 2} M_{AB} \phi^A \phi^B,$$
where $\phi^A, \phi^B$ denote all fluctuating fields.}, \cite{Tripathy:2005qp},
\beq
\label{massmatrixappb}
M=E \left(3{D_a D_d \over D}-D_{ad}\right) \otimes {\bf I} +  D_{ab} \otimes (A \sigma^3 - B  \sigma^1).
\eeq
In Appendix A.1 we give the values of the coefficients, $E,A, B$. 
 The mass matrix in eq.(\ref{massmatrixappb}) is written in a tensor product notation.
The labels, $a,d$ take values, $1, \cdots N$. The   ${\bf I}, \sigma_3, \sigma_1$ matrices act in the 
$2\times 2$ space labeled by $(\delta\xi^a, \delta y^a)$, for fixed $a$, while 
   $D_{ab}, D_aD_b$ matrices act in the $N\times N$ space labeled by the indices, $a,b$.

As was analysed in \cite{Tripathy:2005qp}
 the mass matrix has $N+1$ positive eigenvalues \footnote{We are assuming here that
the attractor values  correspond to a non-singular point in the moduli space, for which the 
moduli space metric is non-singular.}  and $N-1$ zero eigenvalues.
The zero eigenvectors correspond to fluctuations of the type, 
$x^a-x_0^a=(\cos\theta + i \sin \theta) z^a$, where the 
  $z^a$'s satisfies the relation, $D_az^a=0$, and $\theta$ is defined by, 
\beq
\label{deftheta}
\cot\theta= {B \over A-E}.
\eeq

For the $D0-D4$ system in particular, $\theta \rightarrow 0$ and the massless modes consist purely 
 of the real parts of 
the fluctuations, $\delta \xi^a$, subject to the constraint, $D_a \delta \xi^a=0$. 

We now turn to some group theory. 
Let $A \in GL(N,R)$,  be a $N\times N$ matrix   which acts on the $x^a$ variables, as
\beq
\label{transxa}
x^a \rightarrow A^a_b x^b.
\eeq
If we  also transform the  charges, and $D_{abc}$, as follows,
\begin{eqnarray}
\label{tcd}
q_a \rightarrow && q_b (A^{-1})^b_a \cr
p^a\rightarrow && A^a_bp^b\cr
D_{abc} \rightarrow && D_{def}(A^{-1})^d_a (A^{-1})^e_b (A^{-1})^f_c,
\end{eqnarray}
 then we see from eq.(\ref{newk}), 
eq.(\ref{spt}) that the Kahler potential and the superpotential and thus $V_{eff}$ are all left invariant. 
This is the central observation  which will aid our discussion of the corrections to the effective potential. 

The STU model is obtained from a consistent truncation of Type IIA on $K3 \times T^2$ (or Heterotic Theory on $T^6$).
It consists of three vector multiplets, $N=3$, and a prepotential:
\beq
\label{pstu}
F=-{X^1 X^2 X^3\over X^0}.
\eeq
This means $D_{123}=-{1\over 6}$ and all  the other non-zero components of $D_{abc}$ are related to this one
 by symmetries. 

\section{The $D0-D4$ System}
We will now consider the $D0-D4$ system in more detail. Our main goal will be to use group theory considerations
and determine the quadratic terms along the massless directions of the effective potential. 
$D2$ brane charge can be included in the analysis in a straightforward manner, but we will not do so here.

The extremum value for non-susy attractor is given by setting $q_a=0$ in eq.(\ref{nonsusyex}) to be,
\beq
\label{ext04}
x_0^a=ip^a \sqrt{-q_0\over D}.
\eeq
 
\subsection{Mass matrix and Group Theory}
The mass matrix was determined in \cite{Tripathy:2005qp} by a direct calculation, and was discussed above. Here we will see that 
group theory allows this calculation to be carried out much more simply. 
The non-supersymmetric extremum is  given by, eq.(\ref{ext04}). 
The most general quadratic fluctuations take the form, 
\begin{eqnarray}
\label{genquadra}
V_{\rm quadr}
&=&  \left( C_1 D_{ab} + C_2 \frac{D_a D_b}{D} \right) \delta\xi^a \delta\xi^b
+ \left( C_3 D_{ab} + C_4 \frac{D_a D_b}{D} \right) \delta y^a \delta y^b  \cr
  &+& \left( C_5 D_{ab} +  C_6  \frac{D_a D_b}{D} \right) \delta\xi^a \delta y^b ~.
\end{eqnarray}
The coefficients $C_i$ can depend on $q_0$, and $ D\equiv D_{abc}p^ap^bp^c,$
 which are the two invariants made out of the charges, under the transformation eq.(\ref{tcd}). 

For the $D0-D4$ case the effective potential has  a symmetry which is useful to bear in mind. 
It   is invariant under the transformation, 
$x^a \leftrightarrow -{\bar x}^a$.
Clearly  the Kahler potential in invariant under this transformation, and therefore so is  the 
metric, $g^{a {\bar b}}$.
  Since the superpotential is quadratic 
in the $x^a$ fields, this transformation takes, $W\rightarrow {\bar W}$, and $\nabla_a \rightarrow -\nabla_{\bar a}$,
leaving the effective potential invariant. Furthermore, since  the extremum value, eq.(\ref{ext04}),
 is purely imaginary, this 
symmetry is unbroken. Now under this symmetry, $\delta \xi^a \rightarrow -\delta \xi^a$  and 
$\delta y^a$ is left invariant. 
 Thus when  expanding $V_{eff}$ around the extremum no term which contains odd powers of $\delta \xi^a$ can appear. 
This means that $C_5,C_6$ must vanish.

We will now obtain the remaining coefficients in eq.(\ref{genquadra})
 by comparing with the STU model. For this purpose it is enough to 
take the 3 $D4$-brane charges   in the STU model to be all equal, $p^a=p, a = 1, \cdots 3$. 
The quadratic terms, eq.(\ref{genquadra}), for the STU model with these charges  then become, 
\begin{eqnarray}
V_{\rm quadr} &=&
- \frac{p}{3} C_1 \left(\delta\xi^1\delta\xi^2 + \delta\xi^2\delta\xi^3 + \delta\xi^3\delta\xi^1 \right)
- \frac{p}{9} C_2 \left(\delta\xi^1 + \delta\xi^2 + \delta\xi^3\right)^2  \cr
&& - \frac{p}{3} C_3 \left(\delta  y^1\delta  y^2 + \delta  y^2\delta  y^3 + \delta  y^3\delta  y^1 \right)
- \frac{p}{9} C_4 \left(\delta  y^1 + \delta  y^2 + \delta  y^3\right)^2 ~.
\label{findcis}
\end{eqnarray}

Now we directly compute the quadratic terms in the STU model. The effective potential is given by, 
\begin{eqnarray}
V_{\rm{eff}} &=&
- \frac{i }{(x^1 - \bar x^1) (x^2 - \bar x^2) (x^3 - \bar x^3)} f(x,\bar x) ~, 
\label{veffstu}
\end{eqnarray}
where the  function $f(x,\bar x)$ is, 
\begin{eqnarray}
f(x,\bar x) &=& \left[ 4 q_0^2
+ 2 p q_0 \left( x^1 x^2 + x^2 x^3 + x^3 x^1 + x^1 \bar x^2
+ x^2 \bar x^3 + x^3 \bar x^1 + \textrm{c.c.}\right) \right. \cr
&+& p^2 \left\{ 4 \left( |x^1x^2|^2 + |x^2x^3|^2 + |x^3x^1|^2 \right)
+ 2 |x^1|^2 (x^2 + \bar x^2) (x^3 + \bar x^3) \right. \cr
& + & \left.\left.  2 |x^2|^2 (x^3 + \bar x^3) (x^1 + \bar x^1)
+ 2 |x^3|^2 (x^1 + \bar x^1) (x^2 + \bar x^2)
\right\}\right]~.
\label{finstu}
\end{eqnarray}
Here `c.c.' denotes the complex conjugation of all the terms inside the parenthesis.  
At the extremum, $x^a=x_0, a= 1, \cdots 3$, where, 
\beq
\label{valstuext}
x_0=i p \sqrt{-q_0 \over D}.
\eeq

Expanding about the extremum,  we get the quadratic terms to be 
\begin{eqnarray}
V_{\rm quadra}&=& {p^2\over |x_0|} \left(  \left(\delta\xi^1 + \delta\xi^2
+ \delta\xi^3 \right)^2  \right. \cr
&+& \left.  \left(\delta y^1 + \delta y^2 + \delta y^3 \right)^2
- 2 \left(\delta  y^1 \delta  y^2 + \delta  y^2 \delta  y^3 + \delta  y^3 \delta  y^1 \right)\right).
\label{vexpand}
\end{eqnarray}
Comparing, eq.(\ref{vexpand}) and eq.(\ref{findcis})  we then get that 
\begin{eqnarray} C_1 = 0 ~,~ C_2 = - 9 \sqrt{-D \over q_0} ~,~ C_3 = 6 \sqrt{-D \over q_0} ~,~
C_4 = - 9 \sqrt{-D \over q_0} ~. \end{eqnarray}
It is easy to see that this is in agreement with the answer obtained in \cite{Tripathy:2005qp} for the D0-D4 case,
 eq.(\ref{massmatrixappb}).

\subsection{Quartic terms}
We see from the calculation above that since $C_1$ vanishes there are $N-1$  massless modes 
for the $D0-D4$ system on a general $CY_3$ ($N=3$ in the STU model). 
These correspond to fluctuations of the real parts, $\delta \xi^a$,  subject to the 
constraint that $D_a \delta  \xi^a =0$. The remaining modes are all heavy with positive mass. 
To determine if the attractor is stable we need to find the leading corrections along the massless
directions. The general symmetry argument discussed above for the $D0-D4$ case tells us that there are no 
cubic terms in the $\delta \xi^a$  fields so  we must go to quartic order.  This makes the 
 resulting calculation  somewhat complicated. 
In particular we will  need to keep terms which are both quartic in the massless  degrees of freedom, and terms which are cubic
involving both the   massive  and massless degrees of freedom. 
 The latter,  after solving for the massive  fields in terms of the massless
 ones, will generate additional terms that are quartic in the massless variables.

To understand this better  consider a simple model with one massive field $\Phi$ and one massless field $\phi$.
The potential around the extremum is
\begin{eqnarray}
\label{toyp}
V = V_0 + \frac{1}{2} M^2 \Phi^2 + \lambda_1 \phi^2 \Phi + \lambda_2 \phi^4  ~.
\end{eqnarray}
Now solving for the massive field in terms of the massless one and substituting back in the potential gives,
a quartic potential in $\phi$ of the form:
\begin{eqnarray}
\label{tp2}
V_{\rm quartic} = \left( \lambda_2 -  \frac{\lambda_1^2}{2 M^2} \right) \phi^4 ~.
\end{eqnarray}
We see that cubic term  in eq.(\ref{toyp}) has given rise to an additional quartic term in eq.(\ref{tp2}). 

In the $D0-D4$  system
the terms which are quartic to begin with in the light fields (analogue of the $\lambda_2 \phi^4$ terms in 
eq.(\ref{toyp}))  were calculated in \cite{Tripathy:2005qp} and are,
\beq
\label{quarta}
V_{\rm quartic1}=-{9\over 2 D} \left({-D \over q_0}\right)^{3\over2} \left(D_{ab}\delta\xi^a\delta\xi^b\right)^2.
\eeq
(More correctly we need to evaluate this term  subject to the constraint that $D_a\delta\xi^a=0$
to get the  quartic terms along the massless directions.) 
However, the  terms  which originate from cubic terms involving the heavy fields were left out in the analysis in
 \cite{Tripathy:2005qp}. 
We turn to determining these next. 

Since the massless fields arise from $\delta \xi^a$, we are interested in cubic terms involving either 
 two $\delta \xi^a$'s  and one $\delta y^a$, or three $\delta \xi^a$'s. However the latter vanish 
due the symmetry which prevents odd powers of $\delta \xi^a$ from appearing. 
The most general terms involving two $\delta \xi^a$'s
and one $\delta y^a$, from group theoretical considerations must take the form, 
\begin{eqnarray}
V_{\rm cubic} =
 \frac{1}{q_0} \left(C_1 D D_{abc} + C_2 D_{ab} D_c + C_3 D_a D_{bc} + C_4 \frac{D_a D_b D_c}{D} \right)
\delta\xi^a\delta\xi^b\delta y^c ~.
\label{veffcubic}
\end{eqnarray}

At first it might seem that other terms can also appear. For example, a term  of the type, 
 $$ D^{ab}D_{abc} D_{de}  \delta\xi^c \delta\xi^d\delta y^e ~, $$
is allowed by the symmetries. 
However in this  term the  $D^{ab}$ tensor
  is fully contracted with $D_{abc}$, and one can see that such a term cannot arise 
when expanding $V_{eff}$.  
 $D^{ab}$ can only appear through $g^{a\bar b}$ in the
potential. But, since $g^{a\bar b}$ appears in the term $g^{a\bar b}
\nabla_a W \overline{\nabla_bW}$ of the potential, and the $D_{abc}$ tensor  would have to arise either from
$D_aW$ or from $\overline{D_bW}$, it cannot  be fully contracted with $D^{ab}$. 
A similar argument also rules out other possible terms from appearing.   
Thus eq.(\ref{veffcubic}) is the most general cubic term containing two massless and one massive
field. 

 Now, to determine the coefficients $C_i$,
we  compare  with STU model. Once again we choose $p^a=p, a=1, \cdots 3$. 
As discussed in Appendix A.2 one finds  that:
\begin{eqnarray}
\label{valcc}
C_1 = 3  ~,~ C_2 = - 9 ~,~ C_3 = 18 ~,~ C_4 = 27 ~.
\end{eqnarray}
 
We can now solve for the massive modes and obtain the quartic terms for the massless fields.
Since the massless directions  correspond to the $\delta \xi^a$ fields,
 subject to the constraint that $D_a\delta \xi^a=0$, we need only keep the first two terms in eq.(\ref{veffcubic}),
 with coefficients proportional to  $C_1, C_2$.
Instead of solving for all the heavy fields we will here only solve for the $\delta y^a$ fields 
in terms of the $\delta \xi^a$ fields. We will then 
need to restrict the fluctuations in $\delta \xi^a$ to satisfy the constraint $D_a \delta\xi^a=0$, to get the final quartic terms along the massless directions.

Setting $D6$-brane charge, $p^0=0$, in eq.(\ref{massmatrixappb})
 we see that the $\delta y^a$ fields have a mass term,
\beq
\label{massy}
V_{mass}={1\over 2} M_{ab} \delta y^a \delta y^b = E \left({3 D_a D_b \over D} - 2 D_{ab}\right) \delta y^a\delta y^b~.
\eeq
As discussed in  Appendix A.2  solving 
for $\delta y^a$ in terms of $\delta \xi^a$ then gives a quartic term,
\begin{eqnarray}
\label{quartb}
V_{\rm quartic2} & = &  -{3\over 8} \left({-D \over q_0}\right)^{3/2}\left( D^{ab}D_{alm}\delta \xi^l\delta \xi^m 
D_{bpq} \delta \xi^p \delta \xi^q\right) \cr
&& + \left({27 \over 8 D}\right) \left({-D \over q_0}\right)^{3/2} \left(D_{lm} \delta \xi^l \delta \xi ^m\right)^2.  
\end{eqnarray}

Combining, eq.(\ref{quarta}), eq.(\ref{quartb}) we then get the full quartic contribution to be,
\begin{eqnarray}
\label{fquart}
V_{\rm quartic} & = & -{3\over 8} \left({-D \over q_0}\right)^{3/2}\left(D^{ab}D_{alm}\delta \xi^l\delta \xi^m 
D_{bpq} \delta \xi^p\delta \xi^q \right)\cr
&& -{9\over 8 D} \left({-D \over q_0}\right)^{3/2} \left(D_{ab} \delta \xi^a \delta \xi^b\right)^2.
\end{eqnarray}

It is important to again emphasise that in the above  expression we must constrain the $\delta x^a$ fields 
to satisfy the constraint $D_a \delta \xi^a=0$,
 in order to get the required quartic contribution along the massless directions. 

It is  also  useful to rewrite eq.(\ref{fquart})  as follows. The metric on the vector 
multiplet moduli space at the extremum,
eq.(\ref{ext04}),  is given by, 
\beq
\label{met}
g_{a {\bar b}}\equiv \partial_a \partial_{\bar b} K |_{x^a=x_0}=-{3 \over  2 D q_0}\left({3\over2}D_aD_b -D D_{ab}\right).
\eeq
Inverting this, we get the relation, 
\beq
\label{dinv}
D^{ab}={3 \over D} p^ap^b + {3 \over 2 q_0} g^{a \bar{b}}.
\eeq
Substituting in eq.(\ref{fquart}) gives,
\beq
\label{fqtwo}
V_{\rm quartic} 
=  {9 \over 4 D}  \left({-D \over q_0}\right)^{3/2}\left[
-\left(D_{lm} \delta \xi^l \delta \xi^m\right)^2 
+ {1\over 4}\left({-D \over q_0}\right)
\left(g^{a {\bar b}} D_{alm} \delta \xi^l \delta \xi^m D_{bpq} \delta \xi^p \delta \xi^q \right)
\right]. 
\eeq
Now for a non-supersymmetric attractor $({-D \over q_0}) >0$, and for a solution where the attractor value is 
non-singular, $g^{a {\bar b}}$ is  non-degenerate with positive eigenvalues, thus we see that the two terms within 
the square brackets come with a relative  opposite sign.   If the net resultant contribution is positive the 
attractor is stable, else it is unstable. In some cases, and we will see an example of this shortly,
 the two terms can cancel against each other identically.  

Let us close this section on the $D0-D4$ system with some more  comments on the STU model. 
In this case $D_{123}=-{1\over 6}$, and all other non-zero components  of $D_{abc}$ are related to it by symmetries. 
Setting all the $p^a$'s equal \footnote{This entails no loss of generality since the $p^a$'s can be bought to this form by rescaling the $x^a$'s.}, $p^a=p, a=1 \cdots 3$, and   evaluating eq.(\ref{fquart}) one gets, 
\beq
\label{qstu}
V_{\rm quartic}={1\over 4 p}[(\delta \xi^1\delta \xi^2 + \delta \xi^2\delta \xi^3+ \delta \xi^1\delta \xi^3)^2
- \{(\delta \xi^1 \delta \xi^2)^2 + (\delta \xi^2 \delta \xi^3)^2 + (\delta \xi^1 \delta \xi^3)^2 \}].
\eeq
Recall that for the quartic terms of the massless fields alone,
 we need to evaluate this expression after imposing the constraint, $D_a \delta\xi^a=0$. For the STU model this takes the form,
\beq
\label{const}
\delta \xi^1 + \delta \xi^2 + \delta \xi^3=0.
\eeq
On imposing this conditions among the $\delta \xi^a$ fields in eq.(\ref{qstu}) one finds that the quartic term
 identically vanishes. 

In fact one can show that the two massless directions for the STU model are exactly flat directions of the 
effective potential. 
Let $x_0^a=\xi_0^a+ i y_0^a$ denote  the critical value for the field $x^a$. Now, solving for the general
 non-susy critical point of the effective action, eq.(\ref{veffstu}),  one finds that $\xi^a, y^a$ must 
satisfy the four equations, 
\begin{eqnarray}
q_0 - p(\xi^{a \ 2} + y^{a \ 2}) & = & 0, a=1,2,3 \cr
q_0\sum_a\xi^a+ p \prod_a \xi^a  & = &  0 
\end{eqnarray}
These four equations admit a 2 real dimensional moduli space of solutions. The moduli space can be 
parametrised by the real parts, $\xi^a$,  subject to the constraint $q_0\sum_a\xi^a+ p \prod_a \xi^a  =  0$.
At the linearised level this constraint takes the form, $\sum_a\xi^a=0$. This agrees with  the constraint, 
$D_a \delta \xi^a=0$,  we found earlier that the massless fields had to satisfy at the quadratic level. 
We will see in the next section that the existence of these  flat directions for the STU model follows
from duality and the existence of  flat directions for the $D0-D6$ system in general.

At we have pointed out earlier, the quartic terms can make the attractor either stable or unstable. In the following
we demonstrate it by considering an explicit example. 
Consider the model \cite{Behrndt:1996he} with a prepotential: 
$F = (a {X^1}^3 - X^1 X^2 X^3)/X^0$. In this case, it is quite straightforward to compute the quartic term. We again take $p^a=p, 1\cdots 3$, then, $D=p^3(a-1)$, and for a 
non-susy attractor to exist, $({-D\over q_0})={p^3(1-a) \over q_0} >0$. Now
denote $\delta\xi^a$, subjected to the constraint $D_a\delta\xi^a = 0$, to be the 
massless directions. We find, on solving for $\delta \xi^3$ in terms of 
$\delta \xi^1,\delta \xi^2$ that,  
\beq
V_{\rm quartic} = {a \over 2(1+ 3 a) }{p^2 \over q_0} \left({-D \over q_0}\right)^{1/2}
\left(\delta\xi^1 - \delta\xi^2\right)^2 \left((3 a - 2) \delta\xi^1 - \delta\xi^2\right)^2 ~.
\eeq
It can be seen that the quartic term diverges for $a=-1/3$. For all other $a$
it has one flat direction and
one other linearly independent direction which becomes stable or unstable 
depending on the values of $q_0,a,p$. For example, if $0<a< 1$ and  
$p,q_0>0$, the attractor is stable, while if $p,q_0<0$, it is unstable. If on the 
other hand,  
 $-{1\over 3} <a<0$, and $p,q_0>0$, the attractor is unstable, while if $p,q_0<0$, it is
stable.  
This example illustrates that the quartic terms  have considerable structure in them,
we will leave a more detailed study  of these terms and their implications for the 
future.

\section{Adding $D6$ Branes}
We now turn  to considering black holes which carry $D6$-brane charge. 
First we consider the $D0-D6$ system and then discuss the case with  $D0-D4-D6$ brane charges. 
\subsection{The $D0-D6$ Attractor}
A black hole with $D0-D6$ brane charges breaks supersymmetry. 
Here we show that the effective potential at the supersymmetry breaking extremum has flat directions.

A  non-susy extremum  for the case with  $D0-D4-D6$ brane charges  was given above in eq.(\ref{extgen}),
eq.(\ref{t1d6}), eq.(\ref{t2d6}). From 
there we can obtain a  solution 
for the $D0-D6$ case by taking a  limit where the  $D4$ brane charge goes to zero.  
Some care must be exercised in taking this limit. 
Let us start with the $D0-D4-D6$ charges chosen  so that we are  in the branch where,  $s/p_0>1$. 
This means $D/q_0<0$. Now we take the limit of vanishing $D4$ brane charge by scaling all the 
$p^a$'s to go to zero at the same rate, i.e. we take $p^a \rightarrow \lambda p^a$ and take the limit as $\lambda 
\rightarrow 0$. In this limit it is easy to see that $t_1 \rightarrow {2\over |p_0|}$, and since the real part, 
$\xi^a=p^a t_1$, and $p^a$ goes to zero, we find that  $\xi^a \rightarrow 0$. On the other hand, 
$t_2 \rightarrow \left({q_0\over -D}\right)^{1\over 3}\left({1\over |p_0|}\right)^{1\over 3}$, 
this means the imaginary part, 
\beq
\label{limya}
y^a=p^a t_2 = p^a \left({q_0\over -D |p_0|}\right)^{1\over 3},
\eeq
 stays finite in this 
limit, since $|D|^{1/3}$ goes to zero   at the same rate as $p^a$ goes to zero. 
From eq.(\ref{limya}) we see that the resulting values for the  $y^a$'s satisfy the equation \footnote{The attractor value for the 
volume of the $CY_3$ in the $D0-D4-D6$ system we start with is proportional to $V\propto -D t_2^3$. Thus 
$D<0$. Since we are also working with charges for which $D/q_0<0$ this means $q_0>0$.} ,  
\beq
\label{modsp}
D_{abc}y^ay^by^c=-|{q_0\over p_0}|. 
\eeq
So we see that by setting the $\xi^a$ fields to zero and choosing any set  of $y^a$'s which satisfies the relation
eq.(\ref{modsp}) we get an extremum of the effective potential for the $D0-D6$ case.  
This means   there is a moduli space of non-susy solutions for the attractor equations in the 
$D0-D6$ case. For a $CY_3$ with $N$ vector multiplets the moduli space is $N-1$ real  dimensional. 

One can also directly analyse the conditions for an extremum of the effective potential in the $0-6$ case. 
This leads to the same result, that any choice of $\xi^a, y^a$ where $\xi^a=0$ and $y^a$ satisfies the constraint,
eq.(\ref{modsp}) extremises the entropy function. Some steps are indicated in Appendix A.3. 
For the specific case of the STU model one can go further and show that these are in fact all the solutions to the 
attractor conditions. 

Expanding around any non-singular point in this moduli space, 
the general analysis of the mass matrix in \cite{Tripathy:2005qp}
 shows that all the $N+1$ massive fields (which are the real fields
$\xi^a$ and one combination of the $y^a$'s) have a positive mass. We have seen above that the 
$N-1$ massless fields correspond to flat directions and thus are not lifted at cubic or higher order 
in the expansion around the critical point.  Thus the solutions we have found are stable attractors. 

One more comment is worth making. For the STU model $N=3$, so there are two exactly flat directions
in this case when the black hole carries $D0-D6$ brane charges. This agrees with the number of flat directions 
we had found for this model in the $D0-D4$ case. The agreement in fact follows from duality. 
The STU model corresponds to 
taking Type IIA on $K3 \times T^2$. The duality group is $O(6,22) \times SL(2)$. 
There is only one duality invariant, the entropy of the black hole \footnote{Since we are dealing with the 
two derivative action we can take the duality groups to be valued in Reals. More generally the duality groups are 
valued in Integers and there are extra invariants.}. This means  a black hole  with $D0-D6$ charges can be turned after duality 
transformation into a $D0-D4$ black hole  with the same  entropy.
Thus duality tells us that the number of flat direction of the effective potential in the two cases  needed to match.
  More generally using duality one can relate the $D0-D6$ black hole 
in the STU model to a $D0-D2-D4-D6$ black hole.
Thus the non-susy extremum of the  effective potential must have two exactly flat directions  in this more general case as well. 

\subsection{The $D0-D4-D6$ Black Hole}
Finally let us consider a black hole which carries $D0-D4-D6$ brane charges. 
From the discussion in section 2 we know that for the non-supersymmetric extremum  
there are $N-1$ massless directions.
In the $D0-D4$ case the effective potential has a symmetry under the exchange, $x^a \leftrightarrow -{\bar x}^a$,
This symmetry is now broken by the terms in the superpotential dependent on the $D6$ brane charge, $p_0$.
Thus there is no direct argument which says that odd powers of the massless fields 
cannot  appear in the expansion about the non-supersymmetric extremum.

We are interested in the higher order corrections along the massless directions in order to decide if the attractor 
is stable. The first correction which is now allowed by the symmetries is cubic in the massless fields. 
 Along the massless directions we can write $x^a-x_0^a= (\cos\theta + i \sin\theta ) \alpha^a$, where 
the angle $\theta$ was defined in eq.(\ref{deftheta}), and the 
$\alpha^a$'s satisfy the constraint, $D_a \alpha^a=0$. 
Group theory
 considerations tell us that the most general cubic term along the massless directions must take the form,
\beq
\label{gencu}
V_{cubic}=C D_{abc} \alpha^a\alpha^b\alpha^c
\eeq
The coefficient $C$ can depend on $q_0, p^0$ and $D=D_{abc}p^ap^bp^c$, which are the three invariants under the 
$GL(N,R)$ transformation, eq.(\ref{transxa}), eq.(\ref{tcd}),  that can be made 
from the charges and the intersection numbers. 
Therefore by   calculating $C$ in the  STU model and expressing the answer in terms of $q_0, p^0$ and $D$,
 we can  obtain the value of the cubic term in general. 

In fact, we already know from the duality argument given at the end of the last subsection that the two massless directions in the $D0-D4-D6$ case for the STU model must be exactly flat and thus no cubic term can appear in the STU model. This means that $C$ must identically vanish as a function of $q_0, p^0, D$,  and thus there will be no cubic
term for the case of a general Calabi-Yau compactification.

For good measure we have checked this conclusion by directly  calculating the cubic term in the 
STU model \footnote{This calculation was carried out using Mathematica. }. We have found that the cubic term does 
indeed vanish. We have also carried out an analytic calculation to first order in $p^0$ and found that once again 
the cubic term vanishes. 

These considerations correct the earlier results reported in \cite{Tripathy:2005qp}
 where it was stated that the cubic term  is in fact 
non-vanishing.

Since the cubic term vanishes one must now go to the quartic order. Once again group theory tells us that 
 only two terms can appear. These have the same tensor structure as in the $D0-D4$ case eq.(\ref{fquart}),
with the $\delta \xi^a$ fields now being replaced by the $\alpha^a$ fields which (after imposing the constraint
$D_a\alpha^a=0$ ) are the massless directions. The coefficients can be obtained by a comparison with 
 the STU model. 
We know from duality, as has been argued above, that in this case the massless directions are flat, so that no 
quartic term can appear either. This imposes one relation between the two coefficients of the quartic terms. 
The allowed quartic terms then take the form, 
\beq
\label{q046}
V_{quart}=C_1 \left[ D^{ab}(D_{alm}\alpha^m\alpha^n) (D_{bpq}\alpha^p\alpha^q) + {3 \over D} 
(D_{ab}\alpha^a \alpha^b)^2 \right]~.
\eeq
The coefficient $C_1$ can be obtained by a direct calculation in the STU model.  
This calculation is straightforward in principle, but we do not carry it out here and leave it for the future. 
One thing can be said, since we know that the massless directions are 
 exactly flat in the $D0-D6$ system, $C_1$ must vanish 
in the limit when the $D4$ brane charge vanishes, and more generally  when, ${D \over (p^0)^2 q_0} \rightarrow 0$.

\vspace*{1cm}

\noindent
{\Large{\bf{Acknowledgments}}}

\vspace{.2in}
We would like to thank   Rudra Jena for collaboration at an early stage in this work. We are also
grateful to Suresh Govindarajan and especially Atish Dabholkar  for helpful  discussions. P.K.T. is indebted 
to the Institute of Physics, University of Neuchatel for kind hospitality. 
We thank the organisers of the conference, ``ISM06'', held in December 2006 in Puri, 
where some  of this research was done. 
   This research is supported 
by the Government of India. S.P.T. acknowledges support from the Swarnajayanti Fellowship, DST, Govt. 
of India. P.K.T. acknowledges support form the IC\&SR (IITM) Project No. PHY/06-07/17/NFSC/PRAS. 
Most of all we thank the people of India for generously supporting research in String Theory. 

\vspace{.4in}

\noindent
{\Large{\bf{Appendix}}}

\vspace{.3in}

\noindent
{\large{\bf A.1 Some more details }}
\vspace{.2in}

In this appendix we give some more details  regarding the non-susy extrema in the $D0-D4-D6$ case.
These results are taken from \cite{Tripathy:2005qp}. 

In the $D0-D4-D6$ case the non-susy extremum is located at, $x^a=x_0^a=p^a(t_1+ i t_2)$.   
The values of $t_1,t_2$ are determined by the charges. There are in fact two branches for the solution. 
It is useful to define a variable $s >0$ given by,
\beq
\label{defs}
s=\sqrt{(p^0)^2-{4D\over q_0}}.
\eeq
The   two branches correspond to $|s/p^0|<1$ and $|s/p^0|>1$ respectively.
  $t_1$ is given by
\begin{eqnarray}
\label{t1d6}
t_1 = \left\{\matrix{
{2 \over s} {\left(1+{p^0 \over s}\right)^{1/3}-\left(1-{p^0 \over s}\right)^{1/3}\over \left(1+{p^0 \over s}\right)^
{4/3}
+\left(1-{p^0 \over s}\right)^{4/3}} &
|{s\over p^0}|>1 \cr
{2\over p^0} {\left(1-{s\over p^0}\right)^{1/3}+\left(1+{s\over p^0}\right)^{1/3}\over \left(1-{s\over p^0}\right)^{4
/3}+
\left(1+{s \over p^0}\right)^{4/3}} &
|{s\over p^0}| < 1 \cr
}\right.
\end{eqnarray}
and  $t_2$ by:
\begin{eqnarray}
\label{t2d6}
t_2 = \left\{\matrix{
 {4 s \over (s^2-(p^0)^2)^{1/3} \left((s+p^0)^{4/3}+(s-p^0)^{4/3}\right)} &
|{s \over p^0}|>1 \cr
 {4 s \over ((p^0)^2-s^2)^{1/3} \left((|p^0|+s)^{4/3}+(|p^0|-s)^{4/3}\right)} &
|{s \over p^0}|<1 \cr
}\right.
\end{eqnarray}
In these expressions  the branch cuts are chosen so that all fractional powers are real.

The mass matrix for quadratic fluctuations was given in eq.(\ref{massmatrixappb}). 
In this formula,  
\begin{eqnarray}
\label{defeab}
E & = & 12 D e^{K_0} \left(Y_1^2+ {1\over D^2t_2^2} X_1^2\right) \cr
A & =& 12 D e^{K_0} \left({1\over D^2t_2^2}X_2^2-Y_2^2-2Y_1Y_2\right) \cr
B & = & 24  D e^{K_0}  {(X_2-X_1) \over Dt_2}Y_2,
\end{eqnarray}
with, 
\begin{eqnarray}
\label{defx1toy2}
X_1 &=& q_0 + 3 D t_2^2 (1 - p^0 t_1) - D t_1^2 (3 - p^0 t_1) \cr
X_2 &=& q_0 -  D t_2^2 (1 - p^0 t_1) - D t_1^2 (3 - p^0 t_1) \cr
Y_1 &=& - p^0 t_2^2 - 3 t_1 (2 - p^0 t_1) \cr
Y_2 &=& - p^0 t_2^2 +  t_1 (2 - p^0 t_1).
\end{eqnarray}
Here $K_0$ is the Kahler potential evaluated on the solution, eq.(\ref{newk}), and $t_1,t_2$ are as given 
above.  

\vspace{.3in}
\noindent
{\large{\bf A.2~~Determining The Cubic and Quartic Terms}}
\vspace{.2in}

In this appendix we give some more details of the steps 
 leading to the determination of the quartic terms as discussed in section 3.2

First we begin with cubic term involving two massless and one massive field. 
The general structure of such terms is given in eq.(\ref{veffcubic}). 
Evaluating this expression for the STU model, with $p^a=p$,  we get, 
\begin{eqnarray}
\label{cstu}
V^{\rm STU}_{\rm cubic}& =& \left({C_1 p^3 \over 3 q_0}\right)
(\delta y^1 \delta \xi^2 \delta \xi^3 + \delta y^2 \delta \xi^1 \delta 
\xi^3+\delta y^3 \delta \xi^1 \delta \xi^2) \cr
&+& \left({C_2 p^3 \over 9 q_0}\right)(\delta \xi^1\delta \xi^2+ \delta \xi^1\delta \xi^3 
+ \delta \xi^2 \delta \xi^3) \left(\delta y^1+\delta y^2+ \delta y^3\right) \cr
&+& \left({C_3 p^3 \over 18 q_0}\right) ( \delta y^1 (\delta \xi^2 + \delta \xi^3) + \delta y^2 (\delta \xi^1 + \delta \xi^3)
+ \delta y^3 (\delta \xi^2 + \delta \xi^1)) (\delta \xi^1+\delta \xi^2+\delta \xi^3) \cr
&-&\left({C_4 p^3\over 27 q_0}\right)(\delta y^1+\delta y^2+\delta y^3) (\delta \xi^1+ \delta \xi^2+\delta \xi^3)^2.
\end{eqnarray}

The effective potential for the STU model was given in eq.(\ref{veffstu}) and eq.(\ref{finstu}).
 Expanding this directly gives, 
\begin{eqnarray}
\label{cstu2}
V^{\rm STU}_{\rm cubic} 
& = & \left({p^3 \over 2 q_0}\right)\left[-4 \{\delta y^1 \delta \xi^1 (\delta \xi^2 + \delta \xi^3)
  +\delta y^2 \delta \xi^2(\delta \xi^1 + \delta \xi^3) + \delta y^3 \delta x^3  (\delta \xi^1 + \delta \xi^2) \} \right. \cr
& - & \left. 2 \{\delta y^1(\delta \xi^1)^2 +\delta y^2(\delta \xi^2)^2 + \delta y^3 (\delta \xi^3)^2\}\right]~.
\end{eqnarray} 
Equating coefficients, gives the result, eq.(\ref{valcc}). 

To obtain the quartic terms we need to solve for the $\delta y^a$ fields in terms of the $\delta \xi^a$ fields. 
From the cubic terms,   
\beq
\label{relcubic}
V={1\over q_0}\left(C_1 D D_{abc}  + C_2 D_{ab} D_c\right)
\delta y^a \delta \xi^b \delta \xi^c~,
\eeq
and the mass terms, eq.(\ref{massy}),  we get that,
\beq
\label{valheavy}
\delta y^a=- M^{ab} \left[{C_1 D \over q_0} D_{bcd} \delta \xi ^c \delta \xi ^d  + {C_2 \over q_0} D_b D_{cd}
\delta \xi^c \delta \xi ^d\right] ~.
\eeq
Here $M^{ab}$ is the inverse of the mass matrix, eq.(\ref{massy}), and is given by, 
\beq
\label{inm}
M^{ab}={1\over 2 E}\left({3\over D} p^ap^b - D^{ab}\right) ~.
\eeq
Substituting eq.(\ref{valheavy}) for $\delta y^a$ in the cubic terms, eq.(\ref{relcubic}), 
then gives the contribution to the quartic term for the light fields, eq.(\ref{quartb}). 

\vspace{.3in}
\noindent
{\large{\bf A.3~~The $D0-D6$ System}}
\vspace{.2in}

Here we present some more details in the analysis  for the $D0-D6$ case, showing that there is a moduli space
of solutions to the extremum conditions of the effective potential. 
The superpotential in this case is given by, 
\beq
\label{sp06}
W=q_0 + p_0 D_{abc}x^ax^bx^c
\eeq
For the STU model it is easy to see that if $x_0^a$ is a solution to the attractor equations,
then so is $\lambda ^a x_0^a$, where the $\lambda^a$'s satisfy the condition, 
$\lambda^1\lambda^2\lambda^3=1$.
Using this fact we can set the three $x^a_0$'s to be equal, $x^a=x_0, a= 1 \cdots 3$. 
Putting this ansatz into the  effective potential and solving for $x_0$ one finds that the only solution is of the 
form, $x_0=iy$, with, $y^3=|{q_0 \over p^0}|$. More generally then a solution to the attractor conditions takes the form,  
\beq
\label{sla}
D_{abc} y^ay^by^c \equiv -y^1y^2y^3 = -|{q_0 \over p^0}|.
\eeq
  
For a general $CY_3$ we have 
\begin{equation}
\begin{array}{ccc}
V_{eff} & = & e^{K}\left|W\right|^{2}\left[\frac{M}{6}\left(M^{ab}-\frac{3(x^{a}-\overline{x^{a}
})(x^{b}-\overline{x^{b}})}{M}\right)\left(-\frac{3M_{a}}{M}+\frac{\partial_{a}W}{W}\right)\left(\frac{3M_{b}}{M}
+\overline{\frac{\partial_{b}W}{\overline{W}}}\right)+1\right], 
\end{array}\label{genexpr}\end{equation}
 where,
\begin{equation}
\begin{array}{ccc}
M_{ab} & = & D_{abc}(x^{c}-\overline{x^{c}})\\
M_{a} & = & D_{abc}(x^{c}-\overline{x^{c}})(x^{b}-\overline{x^{b}})\\
M & = & D_{abc}(x^{a}-\overline{x^{a}})(x^{b}-\overline{x^{b}})(x^{c}-\overline{x^{c}})\\
g_{a\overline{b}} & = & \frac{3}{M}\left(2M_{ab}-\frac{3}{M}M_{a}M_{b}\right)\\
g^{a\overline{b}} & = & \frac{M}{6}\left(M^{ab}-\frac{3(x^{a}-\overline{x^{a}})(x^{b}-\overline{
x^{b}})}{M}\right) ~. \end{array}\label{usefuldef}\end{equation}

Setting the real parts to zero, $x^a=iy^a$ we now look for a solution to the extremum of the effective potential,
of the form,
\beq
\label{exf}
D_{abc} y^a y^b y^c = C.
\eeq
We find that this ansatz satisfies the equations of motion if 
\beq 
\label{valc}
C=-|{q_0 \over p^0}|.
\eeq

\end{document}